\documentclass[twocolumn]{aastex631}
\usepackage{commath}
\usepackage{floatrow}
\usepackage{soul}
\usepackage{CJK}
\usepackage{threeparttable}
\usepackage{soul}
\usepackage{ulem}
\usepackage[caption=false]{subfig}
\usepackage{appendix}
\usepackage{xspace} 
\usepackage{overpic}

\newcommand{\Ha}{\textrm{H}\ensuremath{\alpha}\xspace}
\newcommand{\Hb}{\textrm{H}\ensuremath{\beta}\xspace}

\newcommand{\HII}{\textrm{H}\textsc{ii}\xspace}

\newcommand{\OIII}{[\textrm{O}~\textsc{iii}]\xspace}

\newcommand{\NII}{[\textrm{N}~\textsc{ii}]\xspace}

\shorttitle{}
\shortauthors{Ren et al.}
\graphicspath{{./}{figures/}}
\begin{document}
\begin{CJK*}{UTF8}{gbsn}

\title{Early Results from GLASS-JWST. XXVI. Spatially Resolved Star Formation and Balmer Decrements at $1.1<z<2.3$ from NIRISS Slitless Spectroscopy\footnote{Based on observations acquired by the JWST under the ERS program ID 1324 (PI T. Treu)}}

\correspondingauthor{Xin Wang}
\email{xwang@ucas.ac.cn}

\author{Pengfei Ren}
\affil{School of Astronomy and Space Science, University of Chinese Academy of Sciences (UCAS), Beijing 100049, China}

\author[0000-0002-9373-3865]{Xin Wang}
\affil{School of Astronomy and Space Science, University of Chinese Academy of Sciences (UCAS), Beijing 100049, China}
\affil{Institute for Frontiers in Astronomy and Astrophysics, Beijing Normal University,  Beijing 102206, China}
\affil{National Astronomical Observatories, Chinese Academy of Sciences, Beijing 100101, China}

\author[0009-0005-3823-9302]{Yuxuan Pang}
\affiliation{School of Astronomy and Space Science, University of Chinese Academy of Sciences (UCAS), Beijing 100049, China}

\author[0000-0002-5815-2387]{Mengting Ju}
\affiliation{School of Astronomy and Space Science, University of Chinese Academy of Sciences (UCAS), Beijing 100049, China}

\author[0000-0001-5860-3419]{Tucker Jones}
\affiliation{Department of Physics and Astronomy, University of California Davis, 1 Shields Avenue, Davis, CA 95616, USA}

\author[0000-0002-3254-9044]{Karl Glazebrook}
\affiliation{Centre for Astrophysics and Supercomputing, Swinburne University of Technology, PO Box 218, Hawthorn, VIC 3122, Australia}

\author[0000-0001-6919-1237]{Matthew A. Malkan}
\affiliation{Department of Physics and Astronomy, University of California, Los Angeles, 430 Portola Plaza, Los Angeles, CA 90095, USA}


\author[0000-0003-2680-005X]{Gabriel Brammer}
\affiliation{Cosmic Dawn Center (DAWN), Denmark}
\affiliation{Niels Bohr Institute, University of Copenhagen, Jagtvej 128, DK-2200 Copenhagen N, Denmark}

\author[0000-0002-6338-7295]{Victoria Strait}
\affiliation{Cosmic Dawn Center (DAWN), Denmark}
\affiliation{Niels Bohr Institute, University of Copenhagen, Jagtvej 128, DK-2200 Copenhagen N, Denmark}


\author[0000-0003-2804-0648]{Themiya Nanayakkara}
\affiliation{Centre for Astrophysics and Supercomputing, Swinburne University of Technology, Hawthorn, VIC 3122, Australia}

\author[0000-0002-8460-0390]{Tommaso Treu}
\affiliation{Department of Physics and Astronomy, University of California, Los Angeles, 430 Portola Plaza, Los Angeles, CA 90095, USA}

\author[0000-0003-0980-1499]{Benedetta Vulcani}
\affiliation{INAF, Osservatorio Astronomico di Padova, Vicolo dell'Osservatorio 5, 35122 Padova, Italy}

\author[0000-0003-3108-0624]{Peter J. Watson}
\affiliation{INAF, Osservatorio Astronomico di Padova, Vicolo dell'Osservatorio 5, 35122 Padova, Italy}

\begin{abstract}

Using JWST/NIRISS slitless spectroscopy, we present spatially resolved Balmer decrement measurements for 79 galaxies at $1.1 < z < 2.3$, which are gravitationally lensed by the foreground cluster Abell 2744.  By stacking \Ha and \Hb emission maps in bins of stellar mass and redshift, we derive radial profiles of nebular dust attenuation and dust-corrected star formation rate (SFR). 
We find tentative evidence that the radial gradients of dust attenuation toward \Ha ($\rm A(\mathrm{H}\alpha)$) vary with both redshift and stellar mass.
At lower redshifts ($z = 1.10$--$1.53$), low-mass galaxies ($\rm  7.0<log(M_*/M_\odot)\leq8.5$) exhibit steeper $\rm A(H\alpha)$ gradients  than higher-mass galaxies ($\rm 9.5<log(M_*/M_\odot)\leq11.0$), while the latter maintain detectable dust attenuation out to larger galactocentric radii.
Galaxies at higher redshifts ($z = 1.76$--$2.29$) show lower attenuation levels. At fixed galactocentric radius, galaxies in the low-redshift bin generally exhibit higher dust attenuation than those at high redshifts, consistent with an increase in dust content toward later cosmic times. Dust-corrected SFR profiles in massive systems at lower redshifts are more spatially extended than those at higher redshifts, consistent with inside-out disk growth at $z\lesssim1.5$.
These results suggest possible differences in attenuation properties across stellar mass and redshift bins, and demonstrate the power of gravitational lensing to probe internal structures in faint galaxies at sub-kiloparsec resolution.

\end{abstract}



\section{Introduction}
\label{sec:intro}

The formation and evolution of galaxies are governed by complex interactions between gas accretion, star formation, stellar feedback, and dust enrichment \citep[e.g., ][]{Larson1992,Bouche2010,Kennicutt2012,nelson2012spatially,Nelson2013,Tacchella2015,nelson2016spatially,Wang2020, Matharu2021, Matharu2022,Liu2025}. Star formation is intimately tied to the molecular gas content and its spatial distribution, and the buildup of stellar mass is modulated by local star formation efficiency, gas inflow and outflow, and dust attenuation \citep[e.g., ][]{Law2009, Kennicutt2012,Madau2014,Vulcani2015,Vulcani2016,Vulcani2017, Wang2019, wangMassMetallicityRelation2022,Inami2022, Reddy2023,Ju2025}. Spatially resolving these components across cosmic time is therefore critical to advancing our understanding of galaxy evolution.

The Balmer lines --- especially the \Ha and \Hb recombination lines --- provide a direct probe of the star-forming ionized gas phase and serve as a robust tracer of the star formation rate (SFR) surface density \citep{kennicutt1998star, calzetti2000dust, Wild2011,Hao2011, Momcheva2013, Price2014, Shapley2022}. However, these optical emission lines are subject to significant attenuation by interstellar dust, especially in dense star-forming regions. Dust grains preferentially absorb and scatter shorter wavelengths, resulting in a suppressed \Hb flux relative to \Ha. The observed flux ratio, known as the Balmer decrement, is sensitive to the line-of-sight dust extinction and is widely used to derive spatially resolved dust attenuation maps \citep[e.g., ][]{calzetti1996reddening, calzetti2000dust,Shivaei2020,Sanders2020}.

Over the past decade, the use of integral field spectroscopy and space-based grism imaging has enabled the first resolved studies of Balmer decrement gradients in distant galaxies. At redshift $z\sim1.4$, \cite{nelson2016spatially} measured spatially resolved Balmer decrement and SFR profiles using HST/WFC3, showing that massive galaxies tend to exhibit centrally peaked dust attenuation and SFR distributions, whereas low-mass galaxies display flatter profiles. These findings support a mass-dependent evolutionary pathway in which more massive systems grow from the inside to outside. Subsequent analyses have revealed correlations between dust attenuation gradients and galaxy morphology, star formation compactness, and stellar mass \citep{Wuyts2013, Hemmati2015}.

The JWST allows spatially resolved studies of rest-frame optical emission lines at $z>1$ with unprecedented sensitivity. In particular, the Near-Infrared Imager and Slitless Spectrograph (NIRISS) offers wide-field slitless spectroscopy that can simultaneously capture \Ha and \Hb across extended galaxy samples \citep{willott2022near, Rigby2023,Glazebrook2023}. \cite{matharu2023first} utilized JWST/NIRISS and NIRCam observations from the CANUCS program to construct stacked Balmer decrement maps for galaxies at $1.0<z<2.4$, finding that low-mass galaxies exhibit steep dust attenuation gradients, while high-mass galaxies show flatter, more uniformly distributed attenuation profiles. These trends were interpreted as evidence of early central dust buildup in massive systems and radially declining SFR distributions, in line with previous studies \citep[e.g., ][]{Tacchella2015,Nelson2023}.

Nevertheless, most such studies have focused on bright, unlensed galaxies due to observational sensitivity limits. Strong gravitational lensing provides a crucial window into intrinsically fainter and lower-mass systems at high redshift \citep[e.g.,][]{Jones2010, Wang2017, Wang2019,Mestric2022,Claeyssens2023,Bergamini2023,Adamo2024,Naidu2024, Messa2025}. Lensing magnification not only enhances the flux of distant galaxies but also improves spatial resolution, enabling sub-kpc-scale analyses of galaxies that are otherwise inaccessible \citep{Jones2010, Livermore2015,Carnall2023, Claeyssens2025}. By combining capabilities of JWST with gravitational lensing, it is now possible to spatially resolve the Balmer decrement in lower-mass, higher-redshift galaxies and directly examine the mass- and redshift-dependence of dust attenuation geometry.

In this work, we use JWST/NIRISS slitless spectroscopy in the field of the strong lensing cluster Abell 2744 to construct stacked \Ha and \Hb emission maps for galaxies. We group galaxies by stellar mass and redshift to investigate trends in radial dust attenuation gradients. We compare our findings to previous results from unlensed fields and explore how lensing-selected samples inform our understanding of dust evolution across cosmic time. Throughout, we adopt the \cite{calzetti2000dust} attenuation law, \cite{chabrier2003galactic} initial mass function and a flat $\Lambda$CDM cosmology with $H_0=70\ \rm km~s^{-1}~Mpc^{-1},\ \Omega_{M} = 0.30,\ \Omega_{\Lambda} = 0.70$.

\section{Observations, Data Reduction, and Sample Selection}\label{sec:DataSamp}

 \begin{figure}
\centering
 \includegraphics[width=1\linewidth]{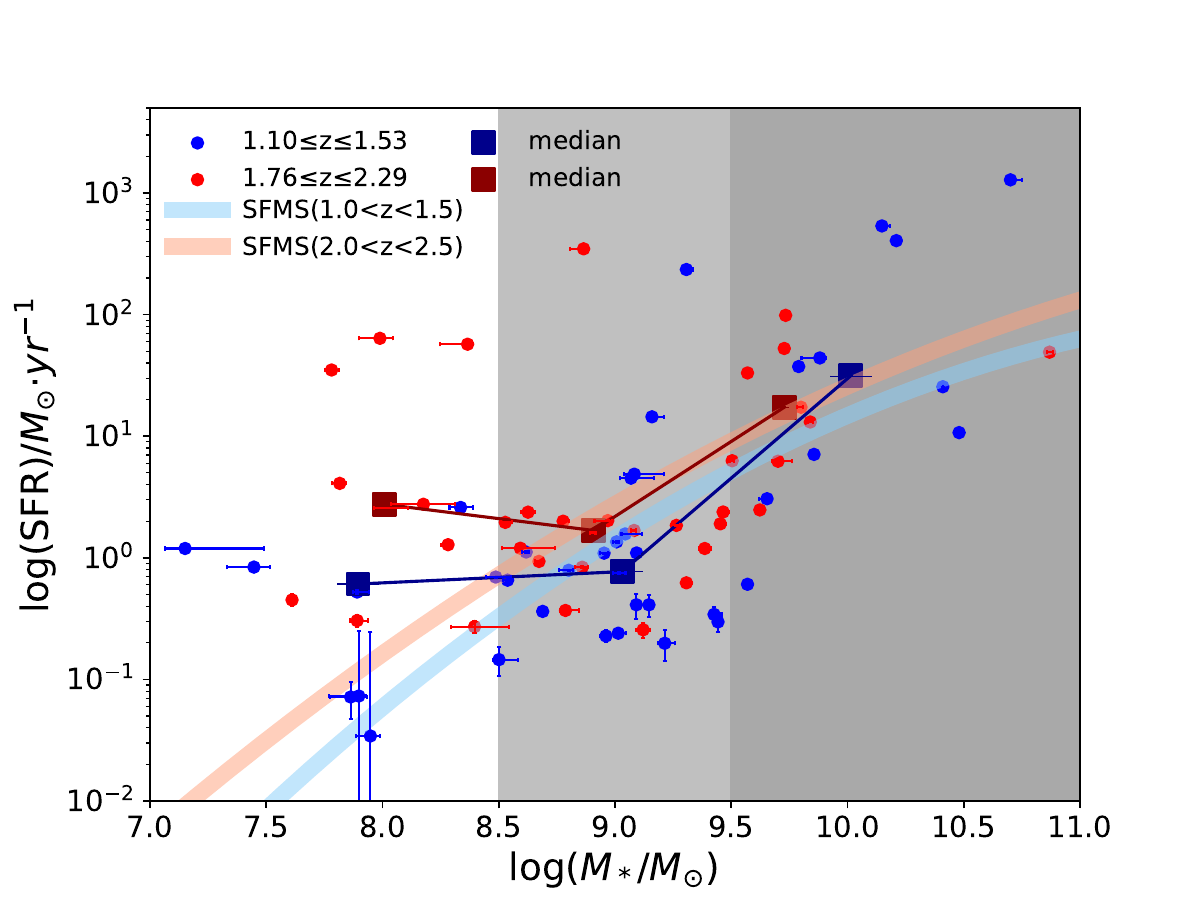}
 \vspace{-1em}
 \caption{SFR and stellar mass ($M_{\ast}$) for our galaxy samples, which are corrected for lensing magnification using the mass model of \cite{Bergamini2023}. Blue and red dots correspond to the galaxies for the $1.10\leq z\leq 1.53$ and $1.76\leq z\leq 2.29$ samples, respectively. The stellar mass ranges adopted for stacking are denoted by the grey shaded regions and medians for each bin are indicated by dark squares. The thick solid lines represent SFMS from \cite{whitaker2014constraining}.
 \label{fig:sample}}
\end{figure}

\begin{figure*}
\centering

\includegraphics[width=1.0\textwidth]{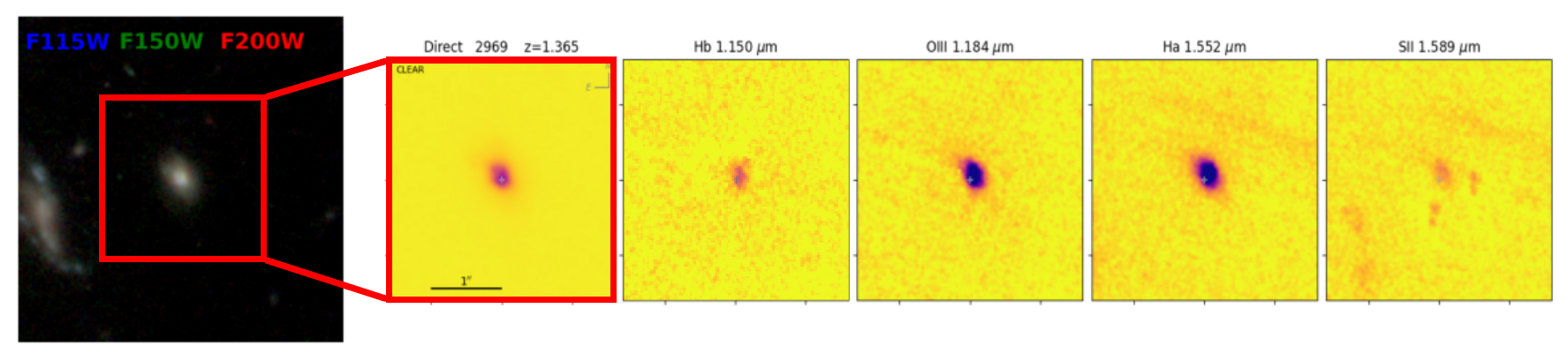}
\includegraphics[width=1.0\textwidth]{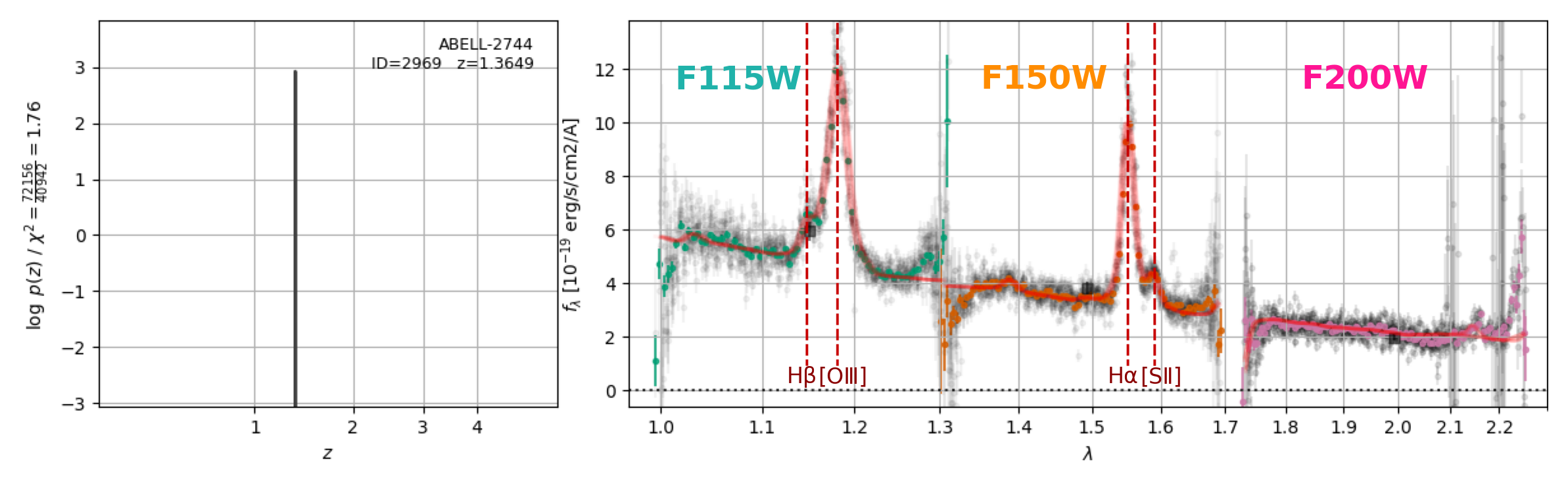}
\caption{Example JWST/NIRISS data products for a galaxy (ID 02969) in the Abell 2744 field. The top row shows the RGB thumbnail, F115W thumbnail and emission-line maps, with observed wavelengths labeled. Each image spans $3^{\prime\prime} \times 3^{\prime\prime}$ ($\sim$25 kpc at $z = 1.36$) at $0\farcs03$ per pixel.
The bottom left panel shows the log p(z) plot,  the unnormalized logarithm of the posterior redshift probability density which is used to assess the relative likelihood of different redshift solutions, showing a sharp peak at $z = 1.36$ with a narrow redshift posterior uncertainty, indicating a well-constrained redshift solution.
The bottom right panel shows the 1D NIRISS grism spectra covered in F115W, F150W and F200W filters.
}

\label{fig:example}
\end{figure*}

We utilize JWST/NIRISS slitless spectroscopy to study galaxies in the gravitationally lensed field of the Abell 2744 galaxy cluster. The spectroscopic data are taken from the GLASS-JWST Early Release Science (ERS) program (DD-ERS-1324; PI: T. Treu), which includes observations of the cluster core ($130\arcsec\ \times\ 130\arcsec$) using the GR150C and GR150R grisms in wide-field slitless spectroscopy mode \citet{treu2022glass}. These observations were conducted between June 28-29, 2022, and July 7, 2023, totaling 18.1 hours of exposure. Additional direct imaging was obtained using three NIRISS filters (F115W, F150W, and F200W) with a combined exposure time of 2.36 hours. This configuration provides low-resolution spectra (R $\sim$ 150) covering the wavelength range $1.0-2.2$ $\mu$m, enabling detection of \Ha and \Hb emission lines for galaxies within $1.1 < z < 2.3$ \citep{acharyya2025spatially,Boyett2022, Wang2022, He2024, watson2025unveiling}.

To estimate stellar masses, we employ the publicly released NIRCam photometric catalog compiled by \cite{paris2023glass}, which integrates observations from GLASS-JWST (PI: Treu), UNCOVER (PIs: Bezanson \& Labb{\'e}), and DDT-2756 (PI: Chen). The catalog includes data in eight NIRCam bands (F090W, F115W, F150W, F200W, F277W, F356W, F410M, F444W)  
and additional HST/ACS and WFC3 bands, spanning a wavelength coverage of $\rm 0.4-5\mu m$. 
This broad spectral baseline samples the rest-frame UV to near-infrared spectral energy distributions (SED) of galaxies and enables robust stellar mass estimation through SED fitting using BAGPIPES \citep[][]{2018MNRAS.480.4379C}. 


In this work, we use the JWST Calibration Pipeline (version 1.19.2) for the basic reduction of the NIRISS data, together with the Grism Redshift and Line Analysis software \texttt{Grizli} \citep[][version 1.12.15]{brammer2022grizli} for subsequent grism modeling and spectral extraction. \texttt{Grizli} adopts a forward-modeling approach to jointly analyze paired direct imaging and grism exposures, producing contamination-subtracted 1D and 2D spectra and best-fit spectroscopic redshifts. Using template 1D spectra and spatial flux distributions derived from the direct images, together with the grism sensitivity and dispersion functions, it generates 2D model spectra. These models are then compared with the observed data via global $\chi^2$ fitting to determine the optimal continuum template combination and Gaussian emission-line amplitudes. During the spectral extraction and modeling process in \texttt{Grizli}, photometric measurements from the reference images are incorporated into the redshift fitting and are also used to constrain the spatial morphology and flux normalization of the spectra. Stellar masses are derived through broadband SED fitting at the spectroscopic redshifts adopted in this work, using photometric measurements based on the same segmentation map as the photometric catalog to ensure consistency.
Our results show excellent consistency with the latest released reductions \citep[e.g.][]{watson2025glass}, further confirming the robustness of our analysis.
We adopt strict criteria to ensure the reliability of spectroscopic redshifts derived from the grism spectra. A secure redshift is required to satisfy the following: (1) reduced chi-squared ($\chi^2_{\rm reduced} < 2$); (2) narrow redshift posterior uncertainty ($\Delta z_{\rm posterior} / (1 + z_{\rm peak}) < 0.005$); and (3) Bayesian Information Criterion (BIC) significantly favoring the spectral fit over a polynomial model that denotes a continuum-only fit (${\rm BIC} > 20$), indicating strong evidence that the redshift solution is driven by genuine spectral features rather than noise fluctuations in the continuum.

To measure Balmer decrements (\Ha/\Hb), we select galaxies with detectable \Ha and \Hb emission lines. Two redshift intervals are defined based on filter coverage: $1.10\leq z\leq 1.53$, where \Ha and \Hb fall into F150W and F115W respectively; and $1.76\leq z\leq 2.29$, with \Ha and \Hb located within F200W and F150W. The median redshifts for these two bins are 1.34 and 2.00, respectively. We further restrict our sample to galaxies with both \Ha and \Hb detected at S/N $>$ 2. We tested removing the S/N criterion in the source selection and find that 28\% of the galaxies have either \Ha\ or \Hb\ detected at S/N $< 2$. Including these systems does not significantly change our results, which remain consistent within the uncertainties across all stellar mass and redshift bins. Given this overall consistency, we adopt the S/N-selected sample in our main analysis to ensure more robust line measurements.


Our analysis is based entirely on star-forming galaxies, and therefore contamination from active galactic nuclei (AGN) must be removed. Owing to the limited spectral resolution of JWST/NIRISS, the \NII and \Ha lines are blended and cannot be separated, preventing the use of the standard BPT diagnostic. Instead, we identify AGN using the mass excitation (MEx) diagram, which relies on the relation between $\OIII \lambda5008/\Hb$ and stellar mass. All of our initially selected galaxies are classified as star-forming systems, with only one galaxy removed from the sample owing to excessively large uncertainties in its emission-line flux ratios.  We cross-matched our sample with the Chandra Source Catalog and found no X-ray detected sources consistent with AGN among our targets. We also visually inspected the grism spectra and found no sources exhibiting broad emission-line features.
Our selection yields a sample of 79 star-forming galaxies across both redshift bins. 

To illustrate the overall properties of the sample, we show the SFR versus stellar mass relation in Figure~\ref{fig:sample}.  We apply the empirical relationships between SFR and \Ha luminosity from \citet{kennicutt1998star} and \citet{chabrier2003galactic}:
\begin{equation}
\rm SFR(H\alpha)=4.68\times{10}^{-42}\frac{L(H\alpha)}{erg\,s^{-1}}[M_{\odot}\,yr^{-1}]
   \label{eq:1}
\end{equation} 

For individual galaxies, the integrated \Ha flux has been corrected using Balmer decrement and lensing magnification model of \cite{Bergamini2023}. Blue and red symbols correspond to the low- and high-redshift bins, and are compared to the star formation main sequence (SFMS) at $1.0 < z < 1.5$ and $2.0 < z < 2.5$ from \cite{whitaker2014constraining}. 
The lower- and higher-redshift samples have median offsets of $-0.02\pm0.01$ and $-0.13\pm0.01$ dex from SFMSs respectively, suggesting that our sample primarily consists of typical star-forming galaxies selected by emission line flux and thus SFR. We also note that galaxies in the lowest-mass bins generally exhibit star formation rates above the SFMS, which may be because these low-mass galaxies with higher specific star formation rates are preferentially detected. Our sample spans a wide mass range ($\rm log( M_*/M_\odot) \sim 7-11$), particularly probing low-mass galaxies enhanced by gravitational lensing magnification. 
These selections provide a well-characterized sample for analyzing spatially resolved Balmer line emission, dust attenuation, and star formation in galaxies across different masses and redshifts.




\section{Measurements}
\label{sec:analysis}

The study of high-redshift galaxies has been hindered by dust obscuration, particularly in the ultraviolet to visible wavelengths, which prevents direct observation of their true characteristics. Dust absorption complicates the accurate assessment of star formation activity and galaxy structure, leading to significant uncertainties in understanding their evolutionary processes. However, by obtaining \Ha and \Hb emission line flux maps, we can derive the flux ratio distribution to calculate the Balmer decrements, providing a quantitative measurement of the dust distribution across different spatial locations. After correcting for dust extinction, we can derive the true SFR distribution of high-redshift galaxies, offering a more accurate foundation for understanding their evolutionary processes.

\subsection{Stacked maps}

\begin{figure}
\begin{overpic}[width=8cm]{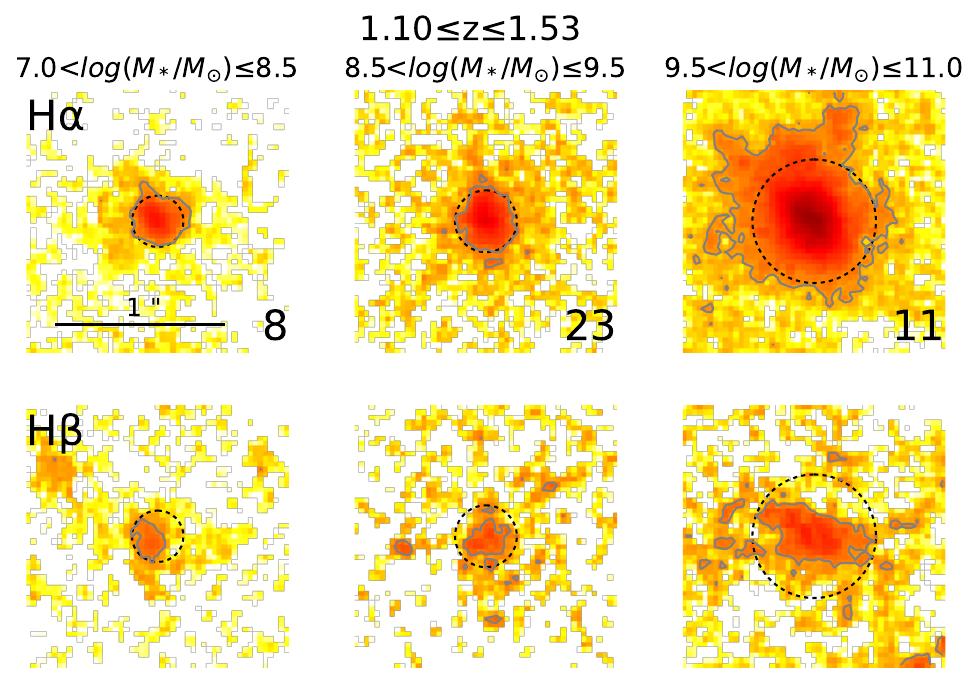} 
\end{overpic}
\begin{overpic}[width=8cm]{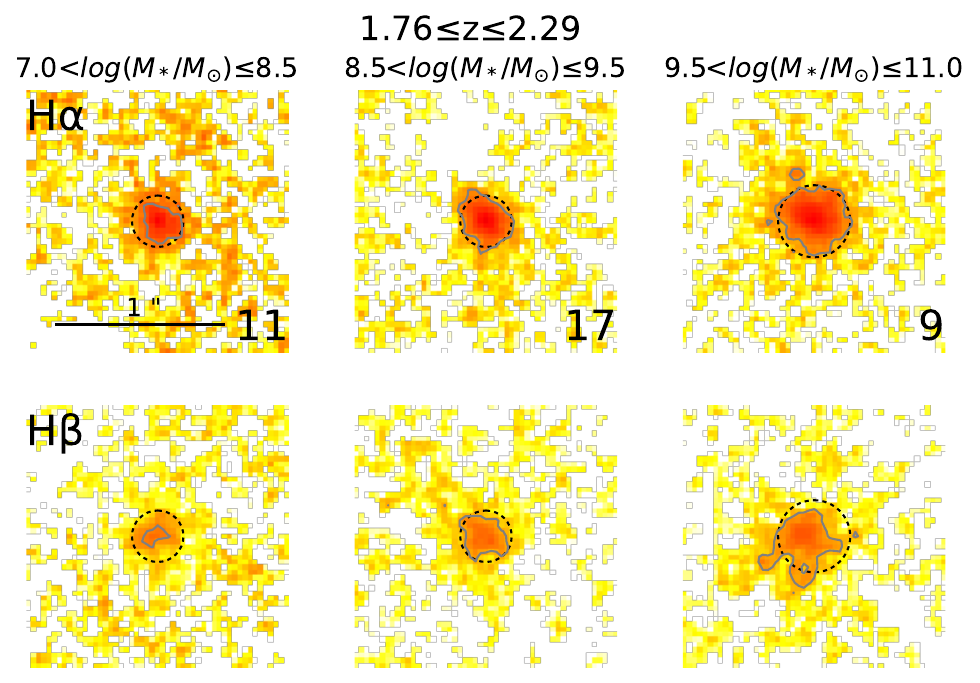} 
\end{overpic}
\begin{overpic}[width=8cm]{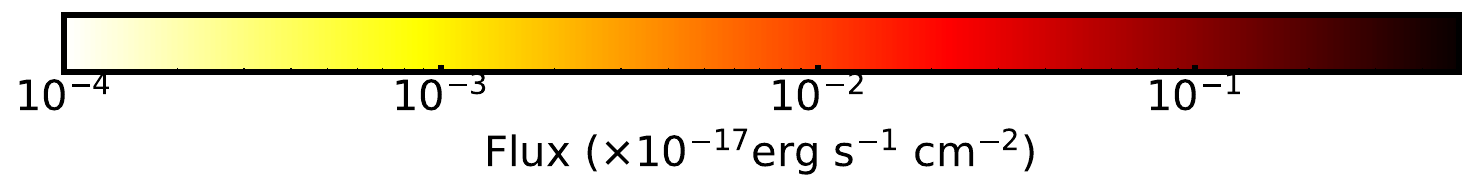}  
\end{overpic}
\vspace{-0.5em}
 \caption{Stacked \Ha and \Hb surface brightness maps. Each composite image spans 50×50 pixels, with a spatial scale of 0.03" per pixel. The number of galaxies included in each stack is indicated in the bottom-right corner of the corresponding \Ha map. The gray line delineates the S/N = 2 boundary, while the dotted circles denote the radial range where the measurement uncertainty of the emission-line profile is less than half of the measured flux.
 \label{fig:map}}
\end{figure}


The median integrated S/N is approximately 20.8 for \Ha and 8.5 for \Hb, suggesting that the data quality of individual galaxy flux maps is insufficient for calculating spatially resolved Balmer decrements.
To significantly enhance the overall S/N and average spatial distributions of \Ha and \Hb, we decide to stack the flux maps. In each of the two redshift bins, galaxies are divided into three mass bins:
$\rm log(M_*/M_\odot) \in(7.0, 8.5]$, $(8.5, 9.5]$, and $(9.5, 11.0]$.
In the low-redshift bin ($1.10\leq z\leq 1.53$), there are 8 galaxies in the lowest stellar mass bin, 23 galaxies in the intermediate mass bin, and 11 galaxies in the highest mass bin, respectively. In the high-redshift bin ($1.76\leq z\leq 2.29$), the sample contains 11 galaxies in the lowest mass bin, 17 in the intermediate mass bin, and 9 in the highest mass bin. These selections define six bins which form the basis of our stacking analysis. 

Before stacking, we applied corrections to the emission-line flux maps of all individual galaxies in the sample, including the $\Ha$-$\NII$ blending correction, the $\Hb$-$\OIII$ blending correction, and the gravitational lensing effect correction. 
Given the spatial and spectral resolution of NIRISS, the blending between \Ha and \NII needs to be carefully considered. To address this issue, we followed the method described in \cite{faisst2018empirical}, which provides the empirical relation between the \NII/\Ha flux ratio, stellar mass, and redshift of galaxies. 
We estimated the expected \NII/\Ha flux ratio using this relation for each of our stellar-mass and redshift bins, and applied the corresponding correction to our measured fluxes. The adopted \NII/\Ha ratios are [0.026, 0.068, 0.185] for low-redshift bin and [0.025, 0.053, 0.112] for high-redshift bin, for the low through high mass stacks respectively.

For the potential blending between \Hb and \OIII, we followed the approach of \cite{matharu2023first} by manually masking the regions that could be affected by imperfect \OIII subtraction. The \Hb radial gradient was measured only within the dotted circles shown in Figure \ref{fig:map}. The outer regions have low S/N and large measurement uncertainties (the error in the measurement is more than half of the measured flux), making reliable analysis there impractical.  While a few individual sources may still show partially overshadowed \Hb emission, we manually masked affected regions for each source before stacking and as a result, the final stacked map represents the sample-averaged properties and further minimizes any residual impact from \Hb-\OIII blending.

Gravitational lensing effects must be taken into account when stacking the emission-line maps. For each individual galaxy, we correct the observed flux maps by its corresponding lensing magnification $\mu$, prior to stacking, so that the intrinsic emission-line surface brightness is preserved. In addition, the gravitational lensing effect also influences the measurement of physical scales, and therefore we performed this correction after stacking. To account for the lensing-induced modification of angular scale, we adopt an effective correction based on the median magnification of the sample.  Under the assumption that lensing conserves surface brightness, the linear dimension is altered approximately by a factor of $\sqrt{\mu}$. Therefore, we convert the pixel scale to a physical scale by dividing the angular scale by $\sqrt{\tilde{\mu}}$ when calculating radial profiles, where $\tilde{\mu}$ is the sample median magnification. Within each stack, the magnification factors span a relatively narrow range, with typical median values of $\sqrt{\tilde{\mu}}$=[$1.32\pm0.11$, $1.45\pm0.09$, $1.27\pm0.06$] for low-redshift and [$1.40\pm0.08$, $1.48\pm0.06$, $1.34\pm0.05$] for high-redshift, such that applying a mean magnification correction after stacking introduces only negligible differences compared to correcting individual galaxies prior to stacking. In addition, because galaxy position angles are randomly oriented, variations in lensing shear average out in the stacking process and are therefore not expected to introduce systematic biases in the recovered radial trends. This provides a consistent physical radial scale for all stacked profiles while retaining proper flux corrections for each galaxy.

To reduce the impact of galaxies with exceptionally high \Ha and \Hb fluxes in the stacked image, a normalization was applied before the stacking process to ensure a consistent flux scale across all sources. 
We adopt the F150W flux as a common normalization reference for all emission-line maps. Specifically, for each galaxy, the H$\alpha$ and H$\beta$ flux maps are normalized by dividing by the total flux measured in the F150W band, which is available for all objects in our sample. Within each redshift bin, the redshift interval is sufficiently narrow that a given emission line falls within a single, fixed filter for all galaxies in that bin, ensuring internal consistency. We then compute the mean of the normalized flux maps. Finally, the stacked map is rescaled by multiplying it by the median F150W flux of the galaxies in the corresponding bin, thereby restoring a consistent absolute flux scale to the stacked results.
The normalization process is expressed as follows:
\begin{equation}
  \overline{f}_{ij} 
  = \frac{1}{N_{\mathrm{obj}}} 
  \sum_{n} 
  \left( \frac{f_{ij}^{n}}{f_{\mathrm{F150W}}^{n}} \right) 
  \cdot f_{\mathrm{F150W}}^{\mathrm{median}}
  \label{eq:2}
\end{equation}
where $f$ represents flux, $f_{ij}^{n}$ denotes the \Ha or \Hb flux of every single pixel ($ij$) of an individual galaxy, ${f}_{F150W}^n$ is the F150W-band flux of the same galaxy, $N_{obj}$ is the number of galaxies in the group, and $f_{F150W}^{median}$ is the median F150W flux of the group.
This weighting would tend to emphasize the brighter galaxies in a group and thus those with the larger stellar masses. However since we have separated the galaxies into separate sub-groups by stellar mass, this bias should be small.

To account for the impact of the point-spread function (PSF) on the stacked emission-line maps, we apply a PSF-correction procedure following the approach described in \cite{szomoru2013stellar} and \cite{nelson2015stars,nelson2016spatially}. For each stacked image, we first construct an empirical PSF model at the median observed wavelength of the corresponding emission line in the given redshift bin, based on the median redshift of the galaxies within that bin. For the \Ha and \Hb measurements, the median observed wavelengths differ from the adopted filter central wavelengths by only $\sim 1\%-3\%$ in both redshift bins. Across the full redshift range of each bin, the corresponding variation in PSF FWHM is generally within ∼10\%. Given that the NIRISS PSF varies only weakly across the wavelength range relevant for \Ha and \Hb at the redshifts considered here, separate PSF models are generated for each line to ensure consistency.
The stacked emission-line maps are then fitted with two-dimensional S{\'e}rsic models using GALFIT \citep{peng2010detailed}, where the model profiles are convolved with the appropriate PSF during the fitting process. After obtaining the best-fit parameters, we generate the intrinsic (unconvolved) S{\'e}rsic model corresponding to the fitted structural parameters. The residual image from the PSF-convolved fit (i.e., the difference between the observed stack and the convolved model) is subsequently added to the unconvolved S{\'e}rsic model. This step preserves non-parametric structures while removing the effects of PSF smearing from the smooth component of the profile.


The stacking procedure is performed by aligning galaxies based on their centroids derived from the F150W broad-band images. This centroid-based alignment is sufficient for measuring azimuthally averaged radial profiles and avoids introducing additional uncertainties associated with position angle measurements, which can be uncertain for low-mass or irregular systems. For completeness, we also test aligning galaxies along their semi-major axes as determined from the continuum images and find that the resulting stacked radial trends are consistent with those obtained using centroid alignment alone. In addition, we assess the impact of the adopted galaxy center by repeating the stacking using the continuum-based centroids rather than the H$\alpha$ peak, and find that the resulting radial profiles remain unchanged within the $1\sigma$ uncertainties.
Figure~\ref{fig:map} presents the stacked \Ha and \Hb emission maps for six galaxy sub-groups. The upper panel shows stacks for three stellar mass groups in the low-$z$ bin, arranged from left to right in order of increasing stellar mass. The lower panel shows stacks for the high-$z$ bin.
In each sub-panel, the top row displays the \Ha flux map and the bottom row shows the corresponding \Hb map. The number of galaxies contributing to each stack is annotated in the bottom-right corner of the \Ha image.

\subsection{Radial Gradients of Dust Attenuation}
\label{sec:attenuation}

\begin{figure*}
    \centering
    \includegraphics[width=1.0\textwidth]{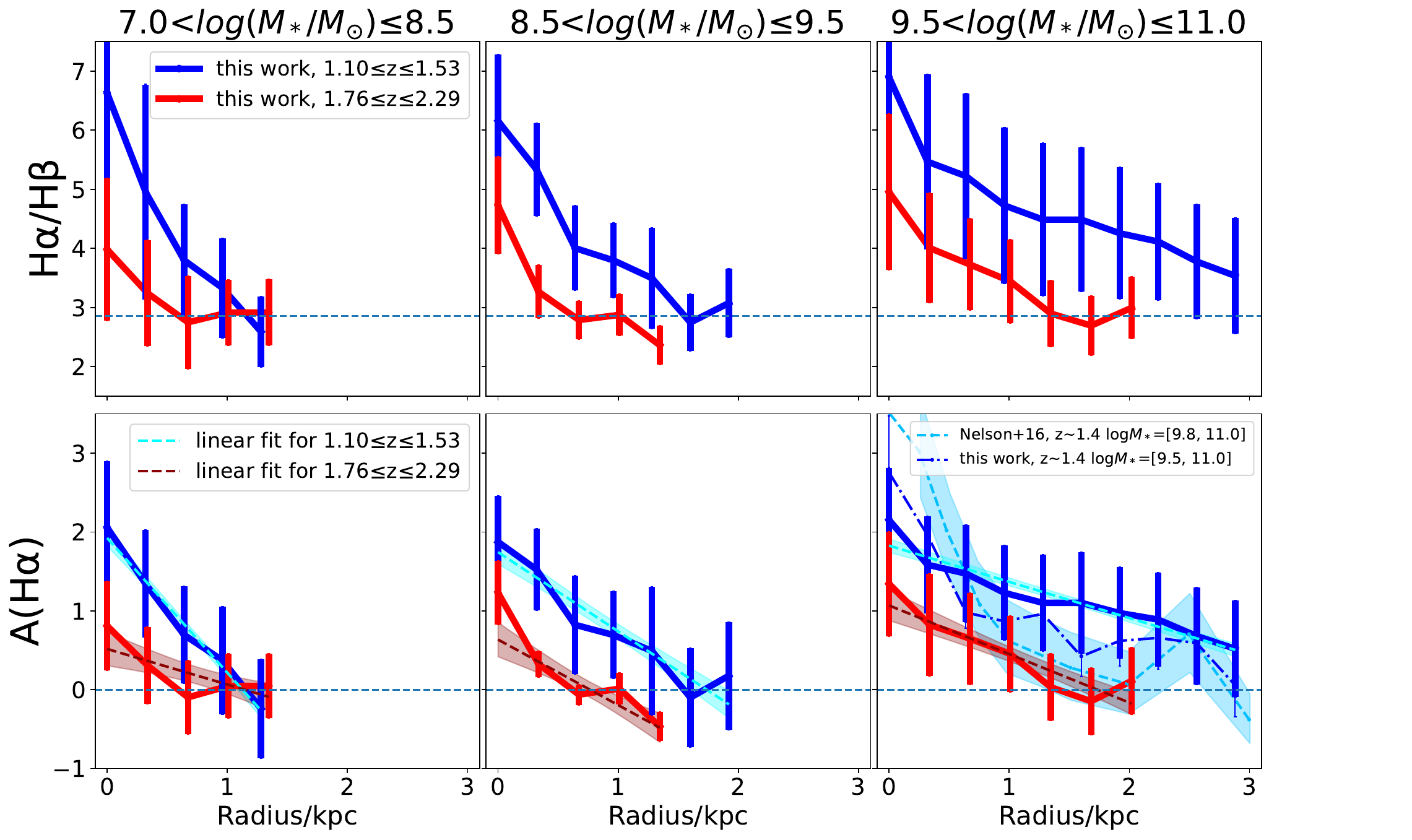}
    \caption{Radial profiles of the Balmer decrement and the dust attenuation towards \Ha emission. The blue and red lines represent galaxies at redshifts of $1.10\leq z\leq 1.53$ and $1.76\leq z\leq 2.29$ respectively. The top panels show the \Ha/\Hb ratios with a dashed line at 2.86 and the bottom row shows dust attenuation towards \Ha using the \citet{calzetti2000dust} dust extinction law. In the bottom row, 
    the light blue dashed line with error envelopes shows result of $z\sim 1.4$ from \cite{nelson2016spatially}.
    }
\label{fig:thiswork}
\end{figure*}

With the \Ha and \Hb stacked flux maps obtained for different redshift and stellar mass bins, we can investigate the spatial distribution of the Balmer decrement to analyze the effects of dust attenuation in high-redshift galaxies. We construct radial profiles of dust attenuation by measuring the Balmer decrement in stacked \Ha and \Hb maps using concentric ring apertures (the top panel of Figure~\ref{fig:thiswork}). Specifically, we apply a circular annuli method with uniform radial intervals of $0\farcs03$ (pixel size of emission maps). These annuli are centered on the \Ha centroid in the stacks and extend out to the radial range where the measurement uncertainty of the emission-line profile is less than half of the measured flux, which corresponds to the dotted circles in Figure~\ref{fig:map}. 

The blue lines in Figure~\ref{fig:thiswork} are the lower-redshift bin, while the red lines correspond to the higher-redshift bin. From left to right, the panels correspond to the lowest, intermediate, and highest mass groups. 


Due to dust extinction, the observed \Ha/\Hb flux ratio deviates from its intrinsic value (\Ha/\Hb)$\rm_{int}$,
which is expected to be 2.86 under Case B recombination at an electron temperature of $T_{\rm e} = 10,000 {\rm K}$ and an electron density of $n_{\rm e} = 100  {\rm cm}^{-3}$ in star-forming regions \citep{Osterbrock2006}. Using the observed \Ha/\Hb ratio, the color excess $\rm E(H\beta-H\alpha)$ can be determined as follows:
\begin{equation}
\rm E(H\beta-H\alpha)=2.5\log\left(\frac{(H\alpha/H\beta)_{obs}}{(H\alpha/H\beta)_{int}}\right)
   \label{eq:3}
\end{equation} 
where $\rm (H\alpha/H\beta)_{obs}$ is the Balmer decrement inferred from observed surface brightness of Balmer lines,

Furthermore, the dust attenuation at the \Ha wavelength, $\rm A(H\alpha)$, can be derived based on the interstellar extinction curve:
\begin{equation}
\rm A(H\alpha)=\frac{E(H\beta-H\alpha)}{k(\lambda_{H\beta})-k(\lambda_{H\alpha})}\times k(\lambda_{H\alpha})
   \label{eq:4}
\end{equation} 
where $\rm k(\lambda_{H\alpha})$ and $\rm k(\lambda_{H\beta})$ mean the values of the adopted reddening curve at the wavelengths of \Ha and \Hb, respectively. We use the dust attenuation law proposed by \citet{calzetti2000dust}, where the expression for $\rm k(\lambda)$ is:
\begin{equation}
\rm k(\lambda)=2.659\times(-1.857+1.040/\lambda)+R_V^\prime
   \label{eq:5}
\end{equation} 
where $R_V^\prime = 4.05\pm0.80$.

To examine the spatial variation of the Balmer decrement, we measure the radial profile of $\rm A(H\alpha)$. 
For each ring, we use Equations~\ref{eq:3} and \ref{eq:4} to derive $\rm A(H\alpha)$ at different radii (the bottom panel of Figure~\ref{fig:thiswork}). This method provides a quantitative assessment of the radial dust attenuation distribution in galaxies. 
The uncertainties on the radial profiles are estimated using a bootstrap resampling of the galaxy sample for each bin and it propagates the statistical variance among galaxies through the full stacking procedure. 
We also compute radial profiles using radial bins with widths comparable to the PSF FWHM ($\sim 0\farcs1$) and the resulting gradients are consistent with those derived at the $0\farcs03$ radial sampling.
These profiles suggest the presence of negative radial gradients at both redshifts, with central regions exhibiting higher dust attenuation than the outskirts.  Massive galaxies appear to show comparatively flatter profiles, consistent with a more uniform dust distribution. We quantify the radial trend by fitting the profiles with the following linear function:

\begin{equation}
\rm A(\mathrm{H}\alpha)(r) = b + c\times\frac{r}{kpc}  
   \label{eq:liner}
\end{equation} 

From low- to high-mass bins, we find $b=[1.92\pm0.10, 1.74\pm0.15,  1.83\pm0.08]$, $c=[-1.71\pm0.12, -1.00\pm0.14, -0.46\pm0.05] $ for low-redshift bins and $b=[0.52\pm0.21, 0.64\pm0.22, 1.07\pm0.19]$, $c=[-0.45\pm0.24, -0.83\pm0.26, -0.62\pm0.14] $ for high-redshift bins.



The overall consistency in gradient shape and normalization supports the universality of dust concentration patterns across independent JWST observations.



\subsection{Profile of Dust-corrected SFR}

\begin{figure*}
\centering
\includegraphics[width=1.0\textwidth,clip,trim={0 0 0 0}]{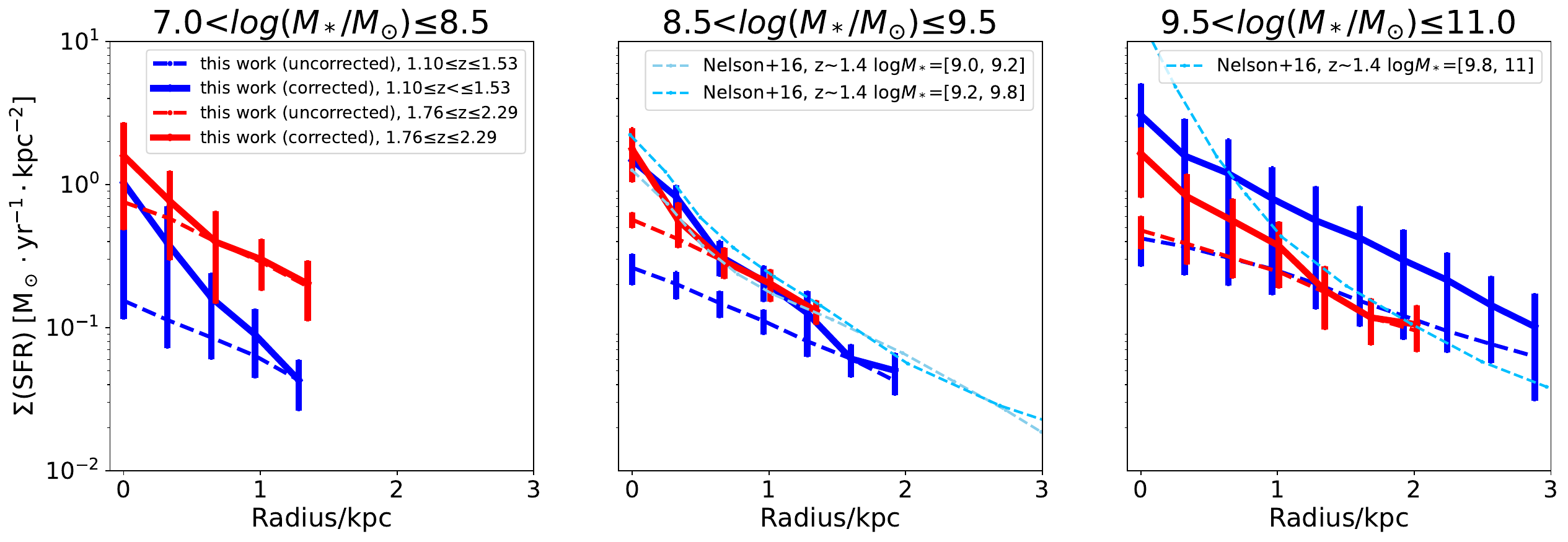}\\
\caption{Dust-corrected SFR radial profiles. The radial distribution of SFR measured directly by \Ha is shown in dashed lines, and the blue and red solid lines represent SFR profiles corrected using the radial Balmer decrement curves for the two redshift bins, respectively. The light blue dashed lines show results of $z\sim 1.4$ from \cite{nelson2016spatially}. The differences between dashed and solid lines imply the detection of previously dust-obscured star-forming regions.
}
\label{fig:SFR_corrected}
\end{figure*}

The Balmer decrements are also correlated with stellar mass and SFR \citep[e.g.,][]{nelson2016spatially, matharu2023first, Shapley2023, Sandles2024}. 
Using Equation~\ref{eq:1}, we first generate the SFR map based on the observed \Ha emission, which does not account for dust attenuation. To account for dust extinction, we utilize the Balmer decrement, a robust indicator of dust attenuation. By measuring the \Ha/\Hb ratio, we recover the intrinsic \Ha luminosity (L(\Ha)), which enables us to derive the dust-corrected SFR map using Equation~\ref{eq:6}. 
\begin{equation}
\rm SFR(dust\ corrected)=SFR(H\alpha)\times \exp\left(\frac{A(H\alpha)}{1.086}\right)  
   \label{eq:6}
\end{equation} 





The radial distribution of the SFR, as shown in Figure~\ref{fig:SFR_corrected}, provides valuable insights into the structural evolution of galaxies. The blue and red lines correspond to those in Figure~\ref{fig:thiswork}, with the dashed lines representing the observed SFR profiles for each redshift group which are not corrected for dust extinction (A(\Ha)). 
In the high-mass sample, the original nearly coincident SFR profiles show divergence after dust correction, with the subsample at $1.10\leq z\leq 1.53$ demonstrating more extended star-forming regions. This observational evidence strongly suggests the detection of previously dust-obscured star-forming activity that has been revealed through our correction procedures.

The regions where the dust-corrected SFR is consistently higher than the observed SFR underscore the substantial effect of dust attenuation on star formation measurements. 
For massive galaxies, those at lower redshift exhibit a more extended SFR profile, indicating ongoing star formation across larger radial scales. These trends offer valuable insights into the mechanisms governing star formation in galaxies of varying masses.


\section{Conclusion and discussion}
\label{sec:discussion}

We have used JWST/NIRISS slitless spectroscopy and NIRCam imaging of galaxies in the gravitationally lensed field of Abell 2744 to measure spatially resolved Balmer decrement and dust attenuation profiles at  $1.10\leq z\leq 1.53$ and $1.76\leq z\leq 2.29$. By stacking \Ha and \Hb emission maps in bins of stellar mass and redshift, we constructed radial profiles of A(\Ha) and SFR.

The redshift-dependent difference in radial dust attenuation suggested by our analysis indicates that low-$z$ galaxies may exhibit somewhat higher $\rm A(\Ha)$ than their high-$z$ counterparts, broadly consistent with recent studies on the evolution of dust content with redshift \citep[e.g.,][]{Shapley2023,jolly2025alma}. 
This pattern has been interpreted as evidence for early dust enrichment and rapid buildup of central star-forming regions at $z>2$, in line with other studies of obscured star formation in massive galaxies \citep[][]{whitaker2014constraining,Reddy2015,Tacchella2015}. These galaxies may still be in early stages of dust production, with lower metallicities and smaller dust-to-gas ratios \citep[e.g., ][]{Shivaei2020,Sommovigo2022}.


We observe possible differences in the dust attenuation profiles of lower-redshift galaxies across stellar mass bins. 
To facilitate a more direct comparison between different mass bins, we quantify the radial attenuation profiles using weighted linear fits as a function of normalized radius ($\rm R/R_e$). We obtain slopes of $-2.19 \pm 0.16$ and $-1.03 \pm 0.11$ $\rm mag/{R_e}$ for low-mass and high-mass galaxies, respectively. 
As for the high-mass galaxies, the flatter A$(\mathrm{H}\alpha)$ profiles may indicate more mature disks with settled dust and gas distributions \citep[][]{ Wang2019}. 
These lower-redshift, high-mass systems may have undergone more substantial outer-disk dust buildup. 
A similar trend is also reflected in the dust-corrected SFR profiles shown in Figure~\ref{fig:SFR_corrected}. These trends are in line with the inside-out growth scenario, where massive galaxies build central stellar mass early and sustain or rejuvenate central star formation, while low-mass galaxies may experience more distributed or bursty star formation histories \citep{nelson2012spatially, nelson2016stars, Tacchella2015}.


At higher redshifts, low-mass galaxies exhibit modest dust attenuation only within a very compact central region ($\sim0.5$ kpc, or $\sim0.35$ $\rm R_e$), while little to no dust attenuation is detected in their outer regions. One possible interpretation is that stellar feedback in these galaxies redistributes or removes dust,  resulting in suppressed attenuation despite ongoing star formation. Recent observational and theoretical studies suggest that bursty star formation is ubiquitous in low-mass galaxies at $z \sim 1{-}3$. Both spectroscopic measurements and broadband SED analyses have shown that the star formation rates of these systems fluctuate on short timescales $(\lesssim 10-50\mathrm{Myr})$, rather than following a smooth, secular history, leading to large temporal variations in their ionizing output and gas content \citep[e.g.][]{looser2023jades,rezaee2023exploring}. Cosmological simulations likewise predict that low–mass galaxies undergo highly time–variable, bursty episodes of star formation, driven by rapid gas inflow and efficient feedback capable of expelling or redistributing the interstellar medium \citep[e.g.][]{sparre2017star,faucher2018model}. This feedback cycle leads to recurrent phases of gas depletion and outflows, which suppress the build-up of interstellar dust\citep{Sandles2024,acharyya2025spatially}.

In several bins --- particularly in the diffuse, low-surface-brightness outskirts of high-redshift systems --- we observe Balmer decrements that fall below the canonical Case B values.
These measurements correspond to regions with relatively low signal-to-noise ratios, where noise may dominate the derived line ratios. Previous high-S/N spectroscopic studies have reported sub-Case B Balmer decrements in high-redshift galaxies and have attributed them to physical effects such as high electron temperatures, density-bounded \HII regions, or complex nebular geometry \citep[e.g.][]{ Sandles2024, Shapley2023}. In future work, we plan to further investigate this issue using larger samples and higher-quality data, enabling more accurate spatially resolved measurements in the outskirts of galaxies. Such observations will allow us to better characterize the spatial distribution of the dust and star formation in high-redshift systems.

The dust attenuation of galaxies at the low-mass end declines rapidly with radius, approaching nearly zero at $r \sim 1 \mathrm{kpc}$. This trend is consistent with the attenuation profiles reported by \cite{matharu2023first} for their lowest-mass bin ($\rm 7.6 \leq log(M_*/M_\odot) < 9.0$).  We also note several discrepancies between our results and those reported in previous studies. For the high-mass regime, the differing trend of the dust attenuation profiles compared to \cite{nelson2016spatially} may partly arise from differences in the adopted stellar mass and redshift ranges, as already suggested by our subsample comparison. In addition, our observations exploit gravitational lensing, which enhances spatial resolution but may also introduce systematic effects in the reconstruction of spatially resolved emission-line maps. The impact of lensing on inferred attenuation gradients warrants further investigation.
Regarding the discrepancies with \cite{matharu2023first}—for example, whether higher- or lower-redshift galaxies exhibit larger average attenuation—the primary driver is likely galaxy-to-galaxy variation in attenuation profiles. Both studies rely on relatively small sample sizes within each mass–redshift bin, which can amplify the influence of intrinsic diversity.
These tensions are expected to be clarified with the increasing number of emission-line galaxy samples from JWST/NIRISS observations in the near future.

Gravitational lensing provides a critical complement to blank-field surveys by enabling resolved measurements in intrinsically smaller, less massive, and less dusty systems. These findings contribute to a more comprehensive picture of how dust and star formation co-evolve across cosmic time. Future work with larger lensing samples and spatially resolved metallicity and gas kinematics will help clarify the co-evolution of dust, star formation, and structural growth in galaxies across redshift.

\section*{Acknowledgements}

We thank the anonymous referee for very constructive comments that help improve the quality of this paper. This paper is dedicated to the memory of our beloved colleague Mario Nonino who passed away prematurely. We miss him and are indebted to him for his countless contributions to the GLASS-JWST project.
This work is supported by the National Key R\&D Program of China No.2025YFF0510603, the National Natural Science Foundation of China (grant 12373009), the CAS Project for Young Scientists in Basic Research Grant No. YSBR-062, the China Manned Space Program with grant no. CMS-CSST-2025-A06, and the Fundamental Research Funds for the Central Universities. XW acknowledges the support by the Xiaomi Young Talents Program, and the work carried out, in part, at the Swinburne University of Technology, sponsored by the ACAMAR visiting fellowship.
TJ acknowledges support from a Chancellor's Fellowship and a Dean's Faculty Fellowship, and from NASA through grant 80NSSC23K1132. 
PW and BV acknowledge support from the INAF Large Grant 2022 ``Extragalactic Surveys with JWST'' (PI Pentericci) and from the European Union – NextGenerationEU RFF M4C2 1.1 PRIN 2022 project 2022ZSL4BL INSIGHT and from the INAF Mini Grant ``1.05.24.07.01 RSN1: Spatially-resolved Near-IR Emission of Intermediate-Redshift Jellyfish Galaxies'' (PI Watson).
This work is based on observations made with the NASA/ESA/CSA JWST, associated with program JWST-ERS-1324.
The JWST data presented in this article were obtained from the Mikulski Archive for Space Telescopes (MAST) at the Space Telescope Science Institute. The specific observations analyzed can be accessed via \dataset[DOI:10.17909/kw3c-n857]{https://doi.org/10.17909/kw3c-n857}.
The MAST at the Space Telescope Science Institute, which is operated by the Association of Universities for Research in Astronomy, Inc., under NASA contract NAS 5-03127 for JWST. 


\bibliographystyle{aasjournal}
\bibliography{refs.bib}

\end{CJK*}
\end{document}